\documentclass[cite,figures]{epl}
\usepackage{graphicx}
\title{The Mott-Hubbard Transition on the $D=\infty$ Bethe Lattice}
\shorttitle{Mott-Hubbard transition}

\author{Claudius Gros, Wolfgang Wenzel, Roser Valent\'\i,
Georg H\"ulsenbeck and Joachim Stolze}

\institute{
Institut f\"ur Physik, Universit\"at Dortmund,
         44221 Dortmund, Germany
          }


\begin{document}
\maketitle

\begin{abstract}
In view of a recent controversy we investigated the Mott-Hubbard transition 
in $D=\infty$ with a novel cluster approach. 
i) We show that any truncated Bethe lattice of order n can be 
mapped exactly to a finite Hubbard-like cluster. 
ii) We evaluate the self-energy numerically for $n=0,1,2$ and compare 
with a series of self-consistent equation-of-motion solutions. 
iii) We find the gap to open continously 
at the critical $U_c\sim 2.5t^\ast (t\equiv t^\ast / \sqrt{4d})$. 
iv) A low-energy theory for the Mott-Hubbard transition 
is developed and relations between critical exponents are presented.
\end{abstract}

{\it Introduction.} --
The Mott-Hubbard (MH) transition, as the metal-insulator transition 
in translationally invariant systems of interacting electrons is called, 
is little understood at present. The nature of the order parameter 
remains unresolved and a number of phenomenological 
scenarios have been proposed to describe the underlying physics. 
In the Brinkmann-Rice \cite{Brinkmann_Rice} scenario the number 
of charge carriers drives the MH transition,
another view \cite{Kohn} suggests a binding/unbinding transition 
of doubly-occupied and empty sites as the appropriate model. 
In this second picture the MH transition
coincides with the convergence radius of a 
large-interaction expansion \cite{annals}.

The study of interacting electrons in the limit of high 
dimensions \cite{Metzner_Vollhardt, Vollhardt_review} 
has proven very useful, as both analytical and numerical methods 
simplify in the limit of infinite dimensions. For instance, 
Monte Carlo studies \cite{QMC_1,QMC_2} are not limited 
by finite-cluster effects, only by the imaginary-time resolution 
in infinite dimensions. Here we focus on the zero-temperature 
half-filled Hubbard model on the infinite-dimensional Bethe lattice 
in the paramagnetic state{\footnote{Note that the Mott-Hubbard 
transition occurs at zero temperature only in the paramagnetic 
sector, while the true ground state is antiferromagnetically 
ordered at half-filling.}}. A Mott-Hubbard transition is 
known \cite{QMC_1,IPT_1} to occur as a function of interaction strength 
and has been examined by Quantum Monte Carlo \cite{QMC_1}, 
by the iterative perturbation theory \cite{QMC_1,IPT_1, IPT_2, IPT_3}, 
by self-consistent diagonalization studies \cite{exact_dia}, 
by a non-crossing approximation \cite{Pruschke_Cox} and a 
modified-equation-of-motion approach \cite{Wermbter_Czycholl}. 
In a recently predicted \cite{IPT_3, exact_dia} scenario 
for the Mott-Hubbard transition the Fermi-liquid effective 
mass would diverge on the metallic side of the transition, 
while no precursor of the transition would be seen on the 
insulating side for any observable, {\emph e.g.} that the 
gap would close discontinuously from a finite value to 
zero at the transition point. Here we want to examine this 
unusual scenario with a novel approach.

Up to date it has not been clear how to formulate a 
systematic exact-diagonalization approach in infinite dimensions, 
as clusters of physical interest contain an infinite number of sites. 
The smallest cluster (see fig. 1) has been shown by 
van Dongen {\emph et al.} \cite{The_Hubbard_Star}, using an 
equation-of-motion approach, to be equivalent to a 
3-site Hubbard cluster. Here we show, for the first time, 
that {\emph any} truncated Bethe lattice may be mapped to 
an effective Hubbard-like cluster. The exact diagonalization 
of these clusters allows to calculate the local Green's function. 
We then estimate the coefficients of the Laurent expansion of the 
self-energy and show that a simple phenomenological low-energy 
theory for the Mott-Hubbard transition can be formulated in terms 
of a truncated Laurent expansion of the self-energy.
 
\begin{figure}[t]
\centerline{\includegraphics[width=0.7\columnwidth,angle=270]{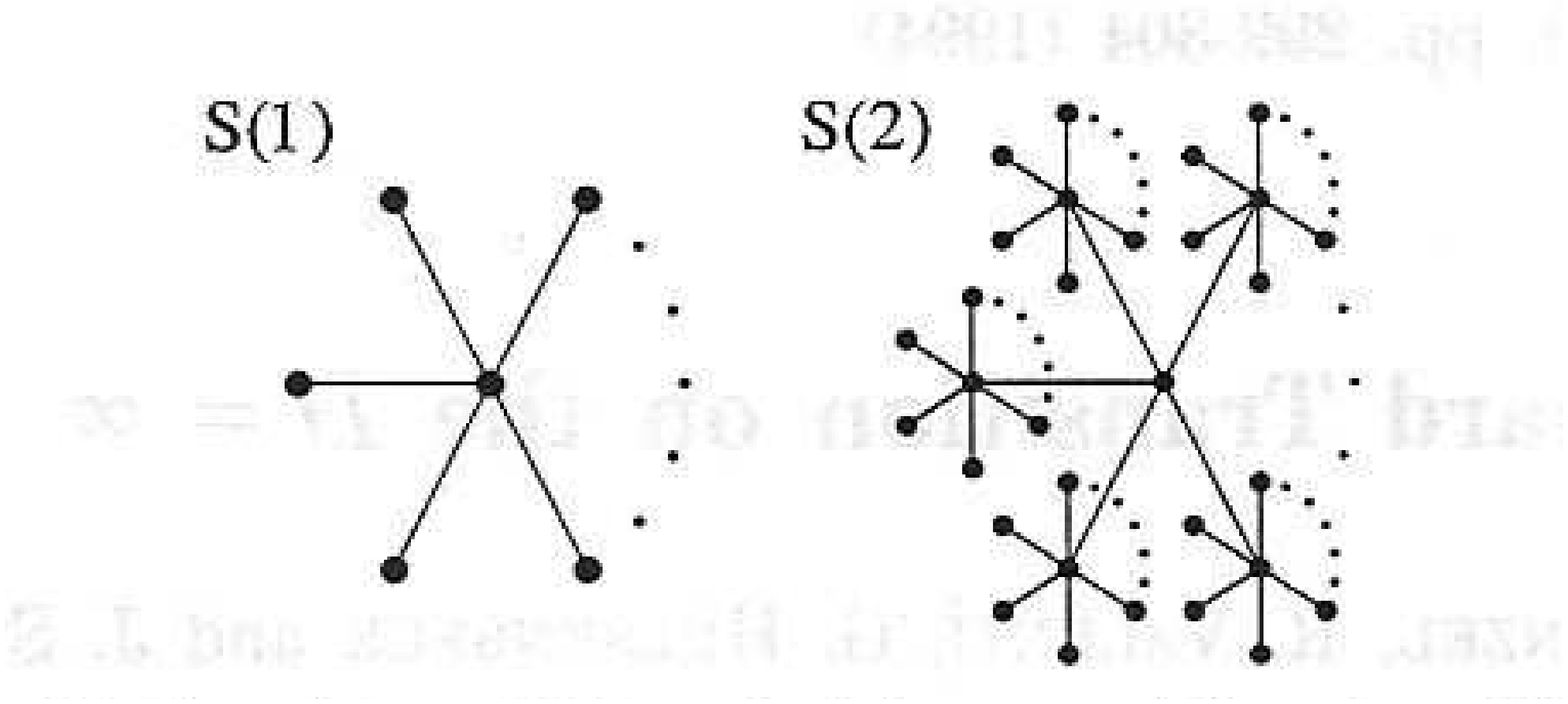}}
\caption{Illustration of the Hubbard star, S(1), and of the
         star of the stars, S(2). The lines connect sites
         connected by hopping matrix elements.}
\label{stars}
\end{figure}
\vspace{3mm}
\par{\it Star of the stars}.-- Any truncated Bethe lattice 
in infinite dimensions can be mapped exactly to a Hubbard-like 
cluster with a finite number of sites. The ground state 
of this cluster can be used to obtain the desired Green's function of 
the truncated lattice. By ``$S(n)$'' we denote a truncated Bethe lattice, 
which contains all sites
 which are separated by at most ``$n$'' steps from the central site.
$S(0)$ denotes the isolated atom, $S(1)$ an atom connected to $2D$ 
n.n. sites and $S(2)$ an atom with $2D$ n.n. and $2D(2D-1)$ 
n.n.n. sites (see fig.\ \ref{stars} for an illustration).
For the Hubbard-model there is an on-site repulsion $U$ on every site and a
hopping \cite{Metzner_Vollhardt} $t=t^*/\sqrt{4D}$ 
matrix element between n.n. sites.

We now consider the zero-temperature, retarded 
Green's functions $G^{(n)}_0(\omega)$ and $G^{(n)}_1(\omega)$ of 
the central site of $S(n)$ and a n.n. site respectively.
In the limit $D\rightarrow\infty$ all diagrams 
contributing to $G^{(n)}_1(\omega)$ which
contain the central site are suppressed as $t^2\sim1/D\rightarrow0$. 
Therefore $G^{(n)}_1(\omega)\equiv G^{(n-1)}_0(\omega)$ and the 
quantity of interest, $G^{(n)}_0(\omega)$ can be calculated recursively.

Denoting by $N_n$ the number of poles of $G^{(n)}_0(\omega)$ we may write
$G^{(n-1)}_0(\omega)=\sum\limits_j^{N_{n-1}} f_j/  
            (\omega+\mu-\varepsilon_j)$, with $f_j\ge0$ and $\sum\limits_j f_j = 1$.
$G^{(n-1)}_0(\omega)$ enters the equation of motion for $G^{(n)}_0(\omega)$
always as $t^2 G^{(n-1)}_0(\omega)$. Therefore $S(n)$ is equivalent to an 
effective system where every one of the $2D$ n.n. sites is replaced by
$N_{n-1}$ non-interacting orbitals which are coupled by hopping matrix 
elements $t_j=t\sqrt{f_j}$ to the central site. The central site Green's 
function couples only to the respective 
symmetric combination of these $N_{n-1}$ orbitals, as they are 
mutually non-interacting. $G^{(n)}_0(\omega)$ is, therefore, 
determined by solving an effective Hubbard cluster where 
one central site (with finite $U$) is connected to $N_{n-1}$ 
other sites (with on-site energies $\varepsilon_j$ and zero $U$)
by hopping matrix elements $t^{(\rm{eff})}_j = \sqrt{2D}\, t_j= t^* \sqrt{f_j/2}$. 
For $N_0=2$ we find $G_0^{(0)}(\omega)= 1/2[1/(\omega-U/2) + 1/(\omega+U/2)]$ at
half-filling and, therefore, $S(1)$ is 
equivalent to an effective 3-site 
Hubbard-cluster \cite{The_Hubbard_Star}. 
The 22 poles of the Hubbard-star Green's function,
$G^{(1)}_0(\omega)$, are reduced by symmetries to an effective $N_1=14$.
Using standard techniques \cite{fraction},  
we have been able to obtain $G^{(2)}_0(\omega)$ 
at half-filling from the ground-state of the 15-site Hubbard-like 
cluster in the parametric sector, which was obtained 
by a direct-diagonalization technique.\\


\par{\it Self-energy}. -- The Laurent series of the self-energy
has the following functional form:
%
\begin{equation}
\Sigma(\omega) = \alpha/\omega+\mu
\label{self_energy} 
               + (1-1/z^*)\,\omega\,
                   \frac{1-\omega\, G_{inc}(\omega)}
                     {1-(1-1/z^*)\,\omega\, G_{inc}(\omega)} ,          
\end{equation}
where $\alpha$ and $z^*$ are parameters depending on $U$. $G_{inc}(\omega)$ 
is an {\it ``incoherent''} Green's function yet to be determined. 
Equation\ (\ref{self_energy}) turns out to be useful in the discussion 
of the low-energy properties of the self-energy.
The Fermi-liquid state corresponds to $\alpha=0$, 
the MH insulating state to $\alpha>0$ \cite{IPT_1,Khurana}.

The first two terms of the r.h.s. of 
eq.\ (\ref{self_energy}) arise from the definition of self-energy 
%
\begin{equation}
\Sigma^{(n)}(\omega)\equiv \omega+\mu
                     - \frac{(t^*)^2}{2} G^{(n-1)}_0(\omega)
                     - \frac{1}{G^{(n)}_0(\omega)} ,
\label{definition}
\end{equation}
on clusters $S(n)$ \cite{cluster_expansion} and particle-hole symmetry at 
half-filling which demands 
$G_0^{(n)} (\omega) = \sum\limits^{N_n/2}_j 2f_j \omega/(\omega^2 
- \epsilon_j^2)$ 
and, therefore, $\alpha = \sum\limits_j^{N_n/2} \varepsilon^2_j/ (2f_j) \geq 0$.

The third term of the r.h.s. of eq. \ (\ref{self_energy}) 
is a consequence of the fact that in infinite dimensions 
the self-energy is site diagonal. 
In the metallic state $(\alpha = 0)$ we write the 
Green's function at the Fermi level as \cite{{Matho}}
 \begin{equation}
 \label{eq_3} 
 \frac{Z}{\omega}+ (1-Z) G_{(\rm{inc})}(\omega) 
\equiv \frac{1}{\omega + \mu - \sum (\omega)} \ .
 \end{equation}
 Solving eq. (\ref{eq_3} ) for $\sum(\omega)$ 
as a function of $G_{(\rm{inc})}(\omega)$ and $Z$, we obtain 
eq. (\ref{self_energy}) with $\alpha = 0$ and $z^\ast \equiv Z$. 
$G_{(\rm{inc})}(\omega)$ is defined through 
eq. (\ref{self_energy}) for the case $\alpha>0$. 
Our numerical results indicate that this $G_{(\rm{inc})}(\omega)$ 
has indeed all the properties of a Green's function, in particular 
that $\Im G(\omega)<0$ and $- \int d \omega \Im G(\omega)/\pi = 1$.
 
\begin{figure}[t]
\centerline{\includegraphics[width=0.7\columnwidth]{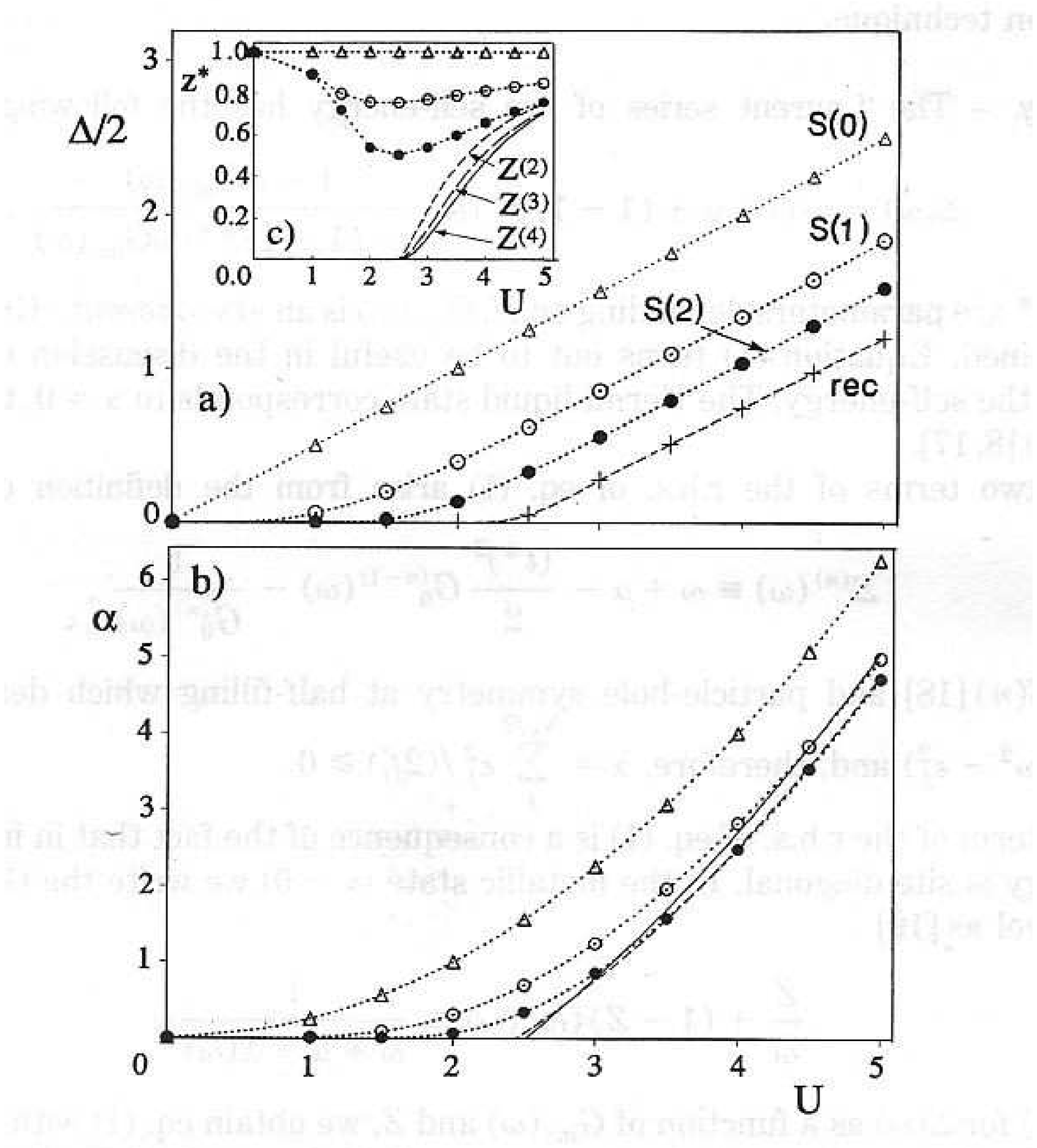}}
\caption{Results from exact diagonalization of the atom ($S(0)$, triangles),
         the star ($S(1)$, open circles) and the star of the stars ($S(2)$, 
         filled circles) as a function of $U$ in units of $t^* $.
         a) The smallest pole of $G^{(n)}_0(\omega)$, which
         corresponds to half the gap. Denoted by `rec' 
         is the prediction for the thermodynamic limit 
         by the recursion method analysis. 
         b), c) The coefficients $\alpha ---(\alpha^
	 {(2)}= \alpha^{(3)}, -\!\!\!-\!\!\!-\!\!\!- \alpha^{(4)}$ and $z^*$ 
         (insert)
         of the Laurent-series of the self-energy:
         $\Sigma(\omega)=\alpha/\omega+U/2+(1-1/z^*)\omega+...$
         as given by  eq.\ (1). 
         Only for $\alpha=0$ is $z^*\equiv Z= m/m^*$. The $\alpha^{(n)}$ and $Z^{(n)}$
         ($n = 2, 3, 4$) are the predictions of the equation-of-motion solutions. $\triangle \ S(1), \circ \  S(2), \bullet \ S(3)$.  
         The dotted lines are guides to the eye.}
\label{gap_alpha_Z}
\end{figure}
~\\[1mm]
\par{\it Results}.--
We have exactly diagonalized the half-filled Hubbard-like
Hamiltonian on S(n) for $n=0,1$ (for $S(2)$ we have $8-\uparrow$ 
and $7- \downarrow$ particles corresponding to $\sim 4 \cdot 10^7$ 
different configurations) and determined the one-particle Green's 
function, $G_0^{(n)} (\omega)$, in the para\-magnetic 
$\big($see footnote ($^1$)$\big)$ superposition of the 
degenerate spin doublets via a continued-fraction expansion \cite{fraction}.

In fig.\ \ref{gap_alpha_Z}a) we have plotted the position of the lowest 
pole in $G^{(n)}_0(\omega)$ (as a function of $U$), corresponding to 
half of the optical gap, $\triangle$. In order to estimate the influence 
of the finite system size on $\triangle$ we have subjected the Green's 
function for $S(2)$ to a continued-fraction analysis which has been 
successfully used \cite{Viswanath} to eliminate finite-size effects 
in numerical calculations of dynamical correlation functions from 
small systems. The first few coefficients of the continued-fraction 
representation of the Green's function allow for reliable estimates 
of its low-energy behavior in the thermodynamic limit. We find that 
the gap vanishes identically for $U \le 2$ in the thermodynamic limit and 
that it is nicely linear for $U \ge 3$, with the same slope as the cluster 
data for $S(2)$. These results, see fig.\ \ref{gap_alpha_Z}a), indicate 
that our estimates of the self-energy parameters from $S(2)$ are 
well controlled and that we may reliably estimate $U_c \sim 2.5 t^\ast$. 

In fig.\ \ref{gap_alpha_Z}b) we present the results for $\alpha$, as defined by
eq.\ (\ref{self_energy}). $\alpha(U)$ increases between $U=2$ and $U=3$ 
by a factor of twelve for S(2), indicative of a gap opening in 
the thermodynamic limit, $S(n\rightarrow\infty)$.

In fig.\ \ref{gap_alpha_Z}c) we show the results for $z^\ast$, as defined by
eq.\ (\ref{self_energy}). Note that $z^\ast$ corresponds 
to the quasi-particle renormalization factor $Z\equiv m/m^\ast$ 
in the thermodynamic limit whenever $\alpha\equiv0$. We find for 
both S(1) and S(2) the minimum of $z^\ast$ occurs at
approximatively the same $U_{min}\sim 2.5t^\ast$ and that $z^\ast$ 
is a continuous function of $U$.

In fig.\ \ref{gap_alpha_Z} we have included the results of a recently 
proposed systematic series of self-consistent equation-of-motion 
solutions \cite{Gros}, $\alpha^{(n)}$ and $Z^{(n)}$ (with $n= 2, 3, 4$). 
The $n$-th-order equation-of-motion solution satisfies the first $n-1$ 
equations of motion exactly, while the $n$-th-order equation of motion 
is decoupled self-consistently \cite{Gros}. $n=2$ corresponds to the 
Hubbard-III solution \cite{Hubbard}. The critical 
$U_c^{(2)} = U_c^{(3)} = \sqrt{6t^\ast} \sim 2.45 t^\ast$ 
and $U_c^{(4)} = \sqrt[4]{40 t^\ast} \sim 2.52 t^\ast$ 
found by the equation-of-motion solutions \cite{Gros} 
agree very well with the $U_c \sim 2,5 t^\ast$ estimated 
from the $S(2)$ cluster with the recursion method, see 
fig. \ \ref{gap_alpha_Z}a) and b), and agrees also 
well with the location of the minimum in $z^\ast (U)$, 
as obtained for both $S(1)$ and $S(2)$, see fig. \ \ref{gap_alpha_Z}c). 
The crowding of lines in fig. \ \ref{gap_alpha_Z}b) for $U>3$ puts 
into evidence the remarkable consistency between the cluster 
results and the equation-of-motion results for $\alpha$.

\begin{figure}[t]
\centerline{\includegraphics[width=0.7\columnwidth,angle=270]{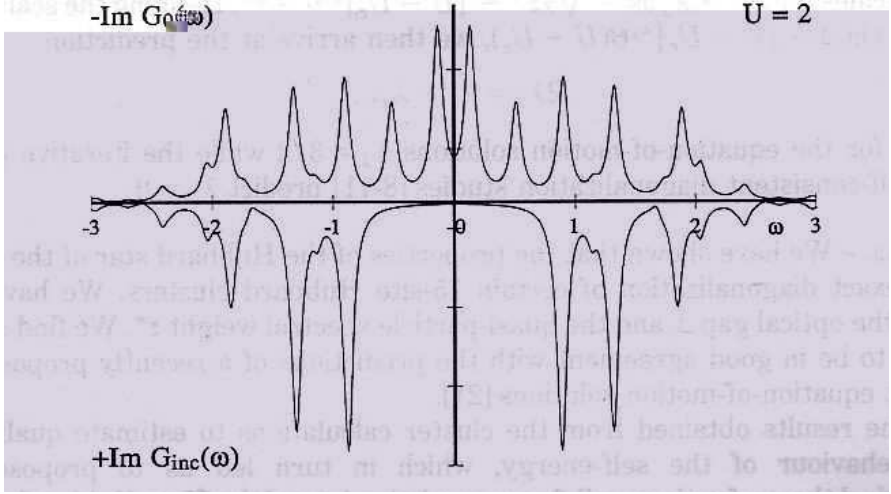}}
\caption{$-Im\, G_0(\omega+i\delta)$ and $+Im\, G_{inc}(\omega+i\delta)$,
         as defined by eq.\ (1), 
         as a function of frequency in units of $t^*$, for $U=2$ and
         $\delta=0.05\, t^*$. Note the
         absence of low-energy poles in the incoherent Green's
         function.}
\label{G_G_inc}
\end{figure}

In Fig.\ \ref{G_G_inc} we plot the results for $-Im\, G_0(\omega+i\delta)$ and
$+Im\, G_{inc}(\omega+i\delta)$ as obtained for the star of the stars for
$\delta=0.05 t^*$ and $U=2$. $G_0(\omega)$ has, as expected, poles at very 
small frequencies, an indication of a gapless density of states in 
the thermodynamic limit. $G_{inc}(\omega)$, on the other hand, has 
no poles at small frequencies, the
structure in $G_{inc}(\omega)$ seems to start at about $U/2$.
In the metallic state the quasi-particle damping if given 
by $\sim\omega^2 Im\, G_{inc}(0)$ and $Im\, G_{inc}(0)$ is 
finite in the thermodynamic limit.
In the insulating state for both $G_0(\omega)$ and $G_{inc}(\omega)$ a gap
$\triangle$ opens and $Im\, G_{inc}(0)=0$.

\vspace{3mm}
\par{\it Discussion}.-- 
Our cluster results for $\triangle, \alpha$ and $z^\ast$ (see fig. 2), the
analysis of the cluster Green's function with the recursion method
\cite{Viswanath} and the series of self-consistent equation-of-motion
solutions \cite{Gros} are all consistent with i) a $U_c \sim 2.5
t^\ast$, ii) the gap $\triangle$ opening continuously at $U_c$ and iii) the
parameter $z^\ast$ of the self-energy representation equation
(\ref{self_energy}) being a continuous function of $U$ at the Mott-Hubbard
transition. These three results do not agree with other studies
\cite{IPT_1}-\cite{exact_dia}, based on the iterative perturbation
theory and self-consistent diagonalization studies, which find {\it a)} a $
U_c - 4.2 t^\ast - 4.7 t^\ast$, {\it b)} the gap $\triangle$ to open
discontinuously at $U_c$ and {\it c)} the parameter $z^\ast$ to be
discontinuous at $U_c$.\\

\par{\it Effective theory}.--
The results for the $G_{inc}(\omega)$ presented in fig. \ref{G_G_inc} indicate 
that $G_{inc}(\omega)$ is a smooth function, for small frequencies, in 
the thermodynamic limit. The first few terms of a small-frequency 
expansion for the self-energy $\big($compare Eq. (\ref{self_energy})$\big)$
\begin{equation}
\sum (\omega) = \frac{\alpha}{\omega} + 
\frac{U}{2}+ \left(1-\frac{1}{z^\ast}\right) \omega + 
i \left(1-\frac{1}{z^\ast}\right) \frac{\gamma}{z^\ast} w^2 \ \ ,
\label{eq.4}
\end{equation}
where $\gamma = G_{inc}(0)$, are then sufficient to describe 
the asymptotic, small-frequency behavior near the Mott-Hubbard transition. 
The three parameters $x \equiv \alpha , z^\ast , \gamma$ entering 
eq. (\ref{eq.4}) are expected to scale like $|U-U_c|^{\lambda_x}$ 
in the critical region $|U-U_c|<<U_c$ of the Mott-Hubbard transition. 
The exact values of the critical exponents, $\lambda_x$, 
are not yet known. The equation-of-motion solutions \cite{Gros} 
predict, for the Bethe lattice $\lambda_x = 1$ and 
$\lambda_{z^\ast} = 2$, the iterative perturbation theory 
and self-consistent diagonalization studies 
\cite{IPT_1}-\cite{exact_dia} $\lambda_{z^\ast} = 2$.  

From the small-frequency expansion of the self-energy, eq. (\ref{eq.4}), 
one can determine the low-frequency behavior of the one-particle 
Green's function $[\omega - (\varepsilon_\beta - \mu) - \sum (\omega)]^{-1}$, 
where the $\varepsilon_\beta$ are the one-particle eigenenergies 
of the Bethe lattice. For the position of the quasi-particle 
poles one finds 
$z^\ast \varepsilon_\beta / 2 [1 \pm 
\sqrt{1 + 4 \alpha / (z^\ast \varepsilon_\beta^2 )}]$. 
For $\alpha > 0$ a Mott-Hubbard gap opens, which 
scales for $\lambda{_z^\ast} > \lambda_\alpha$ as 
$\sim \sqrt{\alpha z^\ast} \sim | U - U_c |^{(\lambda_x + \lambda_{z^\ast})/2}$. 
Defining the scaling exponent of the gap, 
$\triangle$, via $\triangle \sim |U-U_c|^\lambda_\triangle \Theta (U-U_c)_2$ 
we then arrive at the prediction
\begin{equation}
2 \lambda_\triangle = \lambda_x + \lambda_{z^\ast}\ \ .
\label{eq.5}
\end{equation}
We then find for the equation-of-motion solutions $\lambda_\triangle = 3/2$ 
while the iterative perturbation theory and 
self-consistent diagonalization studies \cite{IPT_1}-\cite{exact_dia} 
predict $\lambda_\triangle = 0$.\\

\par{\it Conclusions}.--
We have shown that the properties of the
Hubbard star of the stars can be obtained by exact
diagonalization of certain 15-site Hubbard clusters. We have presented
estimates for the optical gap $\triangle$, and the 
quasi-particle spectral weight $z^*$. We find our 
estimate of $U_c \sim 2.5 t^\ast$ to be in good agreement with 
the predictions of a recently proposed series of self-consistent 
equation-of-motion solutions \cite{Gros}.

We used the results obtained from the cluster calculations to estimate 
qualitatively the low-energy behavior of the self-energy, which in 
turn led us to propose a simple, phenomenological theory for the 
small-frequency behavior of the Mott-Hubbard transition, in 
terms of a truncated Laurent expansion of the self-energy.

We thank {\sc F. Gebhard, G. Kotliar, W. Krauth, T. Pruschke, A.E. Ruckenstein} and {\sc Q. Si} for discussions.
This work was supported by the Deutsche Forschungsgemeinschaft and
by the Minister f\"ur Wissenschaft
und Forschung des Landes Nordrhein-Westfalen.



\end{document}